\documentclass[graybox]{svmult}

\usepackage{type1cm}        % activate if the above 3 fonts are
\usepackage{makeidx}         % allows index generation
\usepackage{graphicx}        % standard LaTeX graphics tool
\usepackage{multicol}        % used for the two-column index
\usepackage[bottom]{footmisc}% places footnotes at page bottom
\usepackage{braket}
\usepackage{newtxtext}       % 
\usepackage[varvw]{newtxmath}       % selects Times Roman as basic font

\makeindex             % used for the subject index
\def\lsim{\lower.5ex\hbox{$\; \buildrel < \over \sim \;$}}
\def\gsim{\lower.5ex\hbox{$\; \buildrel > \over \sim \;$}}

\begin{document}

\title*{Radiative acceleration of relativistic jets from accretion discs around black holes}
\author{Indranil Chattopadhyay\orcidID{0000-0002-2133-9324} \and Raj Kishor Joshi\orcidID{0000-0002-9036-681X} \and Sanjit Debnath\orcidID{0000-0002-9851-8064} \and Priyesh Kumar Tripathi\orcidID{0009-0002-7498-6899} \and Momd Saleem Khan}

\institute{Indranil Chattopadhyay \at Aryabhatta Research Institute of Observational Sciences (ARIES), Manora Peak, Nainital, 263001, India, \email{indra@aries.res.in}
\and Raj Kishor Joshi \at Department of Astronomy, Astrophysics and Space Engineering, Indian Institute of Technology Indore, Khandwa Road, Simrol 453552, India
\and Sanjit Debnath \at Aryabhatta Research Institute of Observational Sciences (ARIES), Manora Peak, Nainital, 263001, India\\
Department of Applied Physics, Mahatma Jyotiba Phule Rohilkhand University, Bareilly, 243006, India
\and Priyesh Kumar Tripathi \at Aryabhatta Research Institute of Observational Sciences (ARIES), Manora Peak, Nainital, 263001, India\\
Department of Applied Physics, Mahatma Jyotiba Phule Rohilkhand University, Bareilly, 243006, India
\and Momd Saleem Khan \at Department of Applied Physics, Mahatma Jyotiba Phule Rohilkhand University, Bareilly, 243006, India}

\authorrunning{Chattopadhyay et al.}
%\institute{Indranil Chattopadhyay \at ARIES, Manora Peak, Nainital \email{indra@aries.res.in}}

\maketitle

% \abstract*{Each chapter should be preceded by an abstract (no more than 200 words) that summarizes the content. The abstract will appear \textit{online} at \url{www.SpringerLink.com} and be available with unrestricted access. This allows unregistered users to read the abstract as a teaser for the complete chapter.
% Please use the 'starred' version of the \texttt{abstract} command for typesetting the text of the online abstracts (cf. source file of this chapter template \texttt{abstract}) and include them with the source files of your manuscript. Use the plain \texttt{abstract} command if the abstract is also to appear in the printed version of the book.}
 \abstract
         {Matter falling onto black holes, {also called} accretion discs, emit intense high-energy radiation. Accretion discs during {hard to hard intermediate} spectral states also emit bipolar outflows. Radiation drag was supposed to impose the upper limit on the terminal speed. It was later shown that a radiation field around an advective accretion disc imposes no upper limit on speed, about a few hundred of Schwarzschild radius from the disc surface. We {study
         radiatively driven electron-proton and electron-positron jets, for gemeotrically thick and slim transonic discs} by using numerical simulation. We show that pair-dominated jets can reach ultra-relativistic speeds by radiation driving. We also discuss at what limits radiative acceleration may fail.}
\vspace{.8cm}

\section{Introduction}
Matter accreting onto black holes emits intense radiation in a {wide} range of the electromagnetic spectrum. Active galactic nuclei and microquasars are supposed to harbour black holes, and they are also observed to be associated with bipolar jets of varying energetics. Jets also originate from the accretion disc around the central black hole. So, it is natural to investigate the possibility of radiative acceleration of ejected matter into relativistic jets. The minimal criterion for radiative acceleration is that the jet should be optically thin. It was, however, soon realised that for optically thin medium and extended sources, i. e., jet medium close to the accretion disc, the radiation field penetrates the medium. So matter plying through this radiation field will experience a decelerating force proportional to various components of the local three-velocity of the jet \cite{sw81,i89,f96}.
This force is called radiation drag. Fukue and his collaborators investigated this aspect of the interaction of disc radiation with the outflow jets both in the fluid as well as in the particle limit
\cite{f99,f05,fth01}. The general consensus on the radiative driving of jets around different disc models is that for certain disc models (such as an infinitely extended Keplerian disc), the jet terminal speed is mildly relativistic \cite{i89}. For other models (like a thick disc or a combination of ADAF and Keplerian disc), the jets are relativistic with a Lorentz factor of about a few \cite{sw81,fth01}. It is accepted that ultra-relativistic jets cannot be produced by radiation driving.
More recently, a number of numerical simulations of super-critical accretion disc showed that outflows driven by disc radiation can reach up to speeds $0.3-0.4c$, where $c$ is the speed of light in vacuum \cite{eg85,sn15,ommk09}.

Parallelly, a group of researchers investigated the issue of radiative driving of outflowing matter by the radiation field around an advective accretion disc \cite{cc00a,cc00b,cc02,cdc04,c05,vkmc15,vc19}.
In these analytical investigations, the authors showed that the radiation field from advective discs does not impose any upper limit on terminal speed. The radiation from advective discs mostly originates from the inner part of the disc, and at a few hundred Schwarzschild radii above, the disc ceases to be an extended object. In such cases, the upper limit imposed on jet speed is the speed of light, in other words, no upper limit on speed. That is, if the disc is sufficiently bright, jets can be accelerated to high speeds.     
Numerical simulations showed that not only can one accelerate jets to high speeds, but for given disc geometry, disc radiation can accelerate pair-dominated subsonic jets to ultra-relativistic speeds \cite{jdc22,jctt24}.  
In this paper, we discuss the basics of radiative acceleration and why this process is more efficient around an advective disc than around other disc models.

In this work we are not studying the generation mechanism of the jet. Advective discs in a significant region of the energy-angular momentum parameter space harbours accretion shocks. Various numerical simulations have shown that thermal gradient force in the post-shock region drives bipolar jets in an orthogonal direction to the equatorial plane of the accretion disc \cite{msc96, mrc96,lckr16,dc14}. In this paper, we assume the jet has been launched by the mechanism of thermal driving of the post-shock disc (PSD), and the jet is accelerated by the radiation of the accretion disc.
The accretion disc is auxiliary and, apart from being the source of radiation, plays no part in the simulations.

In this paper, we solve time-dependent equations of motion of a relativistic gas impacted by the radiation of the accretion disc. We use a TVD {scheme} based on relativistic {hydrodynamics} \cite{rcc06}, but incorporating gravity only through the time-time component of the metric tensor, the so-called `weak field limit' \cite{jctt24}. 
We consider all the ten independent components of the radiation tensor from an advective disc \cite{c05}. The radiation field has two extended sources: (i) the post-shock part of the accretion disc or PSD and (ii) the sub-Keplerian pre-shock disc or SKD. The two sources produce a very interesting radiation field. In this paper, we inject the jets with very low and subsonic velocity and study how much the jets can be accelerated.

In the next section, we will present the assumptions and equations of motion, and then, we will present the results and draw concluding remarks.

\section{Assumptions and equations of motion}
We have incorporated gravity through the weak field approximation in a special relativistic metric written in cylindrical coordinates ($r,~\theta, ~ z$). {If the metric is generally represented as $ds^2=g_{\mu \nu}d^\mu dx^\nu$, where $\mu \equiv(t,~r,~\theta,~z)$, then the weak field assumption} implies, that the gravitation enters through $g_{tt}$ {or, the time-time component of the metric tensor. Therefore,}
\begin{equation}
ds^2=-(1+2\Phi)c^2dt^2+dr^2+r^2d\theta^2+dz^2; \,\,\,\, \Phi=\frac{GM}{R-r_{\rm g}}
\label{eq:metric_potn}
\end{equation}
Here, $R=\sqrt{r^2+z^2}$ is the radial distance from the central object, ($r,\theta,z$) are the usual coordinates in cylindrical geometry, and $r_{\rm g}=2GM/c^2$ is the Schwarzschild radius. The gravitational force is expressed {via a pseudo-Newtonian potential $\Phi$ and is} called Paczy\'nsky-Wiita potential \cite{pw80}.
The equations of motion are $T^{\mu \nu}_{;\nu}=0$, and $(nu^\nu)_{;\nu}=0$, in terms of components they are:

\begin{subequations}
\begin{equation}
\frac{\partial D}{\partial t}+\frac{1}{r}\frac{\partial}{\partial r} \left[r\alpha Dv^r\right]+\frac{\partial}{\partial z}\left[\alpha Dv^z\right]=0    
\label{eq:continuity}
\end{equation}

\begin{equation}
\frac{\partial M^r}{\partial t}+\frac{1}{r}\frac{\partial}{\partial r}  \left[r\alpha\left(M^rv^r+p\right)\right]+\frac{\partial}{\partial z}\left[\alpha M^rv^z\right]=\frac{\alpha p}{r}+\frac{\alpha M^\theta v^{\theta}}{r}-E\frac{\partial \alpha}{\partial r}+G^r    
\label{eq:momentum_r}
\end{equation}

\begin{equation}
\frac{\partial M^\theta}{\partial t}+\frac{1}{r}\frac{\partial}{\partial r}\left[r\alpha M^\theta v^r\right]+\frac{\partial}{\partial z}\left[\alpha M^\theta v^z\right]=-\frac{\alpha M^\theta v^r}{r}+G^\theta    
\label{eq:momentum_theta}
\end{equation}

\begin{equation}
\frac{\partial M^z}{\partial t}+\frac{1}{r}\frac{\partial}{\partial r}\left[ r\alpha M^zv^r\right]+\frac{\partial}{\partial z}\left[\alpha (M^zv^z+p)\right]=-E\frac{\partial \alpha}{\partial z}+G^z    
\label{eq:momentum_z}
\end{equation}

\begin{equation}
\frac{\partial E}{\partial t}+\frac{1}{r}\frac{\partial}{\partial r} \left[r\alpha (E+p)v^r\right]+\frac{\partial}{\partial z}\left[\alpha (E+p)v^z\right]=-M^r\frac{\partial\alpha}{\partial r}-M^z\frac{\partial\alpha}{\partial z}+G^t.   
\label{eq:energy}
\end{equation}
\end{subequations}
In equations, \ref{eq:continuity}-\ref{eq:energy}, $D,\, M^r,\, M^z,\,E$ is the mass density, the radial and axial component of momentum density, and the total energy density of the fluid, respectively. Here $v^r,\,v^\theta,\,v^z$ are the components of three-velocity and $\alpha=\sqrt{(1+2\Phi)}$. It may be noted that the time and space components of 4-velocities are given as
$u^t=\gamma/\sqrt{\alpha}$, $u^i=\gamma v^i/\sqrt{g_{ii}}$, respectively, here $g_{ii}$ are the {space components of the} metric tensor components and $\gamma=1/\sqrt{(1-v_iv^i)}$ is the Lorentz factor. $G^\mu$ are the components of radiation four-force density given as 

\begin{subequations}
\begin{equation}
G^r=G^{\hat{r}}_{co}+\frac{\gamma-1}{v^2}v^rv_iG^{\hat{i}}_{co},
\label{eq:gr}    
\end{equation}
\begin{equation}
G^\theta=\frac{1}{r}\left[G^{\hat{\theta}}_{co}+\frac{\gamma-1}{v^2}v^\theta v_iG^{\hat{i}}_{co}\right],
\label{eq:gtheta}    
\end{equation}
\begin{equation}
G^z=G^{\hat{z}}_{co}+\frac{\gamma-1}{v^2}v^zv_iG^{\hat{i}}_{co},
\label{eq:gz}    
\end{equation}

\begin{equation}
G^t=\frac{\gamma}{\alpha}v_i G^{\hat{i}}_{co},   
\label{eq:gt}
\end{equation}
where,
\begin{equation}
G^{\hat{i}}_{co}={\rho_e} F^i   \,\,\,\,\,\, (i=r,\theta,z),
\label{eq:gico}
\end{equation}
$F^i$ are the components of the flux measured in the observer frame, {and the general form of which is given as},
\begin{align}
F^i &=-\gamma^2v^i\mathcal{E}+\gamma\left[\delta^i_j+\left(\gamma+\frac{\gamma^2}{\gamma+1}\right)v^iv_j\right]\mathcal{F}^j \nonumber \\
&-\gamma v_j\left[\delta^i_k+\frac{\gamma^2}{\gamma+1}v^iv_k\right]\mathcal{P}^{jk}.    
\label{eq:Fico}
\end{align}   
\end{subequations}

In equation \ref{eq:Fico}, $\,\mathcal{E}=\frac{\sigma_{\rm T}}{m_{\rm el} c}E_{\rm rd},\,\mathcal{F}^i=\frac{\sigma_{\rm T}}{m_{\rm el} c}{F}_{\rm rd}^i,\,\mathcal{P}^{jk}=\frac{\sigma_{|rm T}}{m_{\rm el} c}{P}_{\rm rd}^{jk}$. Here, $E_{\rm rd},\, F_{\rm rd}^i,\, P_{\rm rd}^{ij}$, represent the radiation energy density, three components of radiation flux, and six independent components of radiation pressure \cite{c05,vc19,jctt24}. $\rho_e$ is the mass density of leptons and $m_{\rm el}$ is the mass of electron. $\sigma_{\rm T}$ is the Thomson scattering cross-section.

We solve these equations using a TVD {scheme}. For a detailed understanding of the eigenstructure and the code, one may refer to \cite{rcc06}.
The thermodynamics is governed by a relativistic equation of state known as CR EoS\cite{cr09}. {CR EoS is given as
\begin{align}
e&=\rho c^2 f(\Theta, \xi); \mbox{ where, } \\
& f(\Theta,~\xi)= 1+(2-\xi)\Theta\left[\frac{9\Theta+6/\eta}{6\Theta+8/\eta} \right]+\xi\Theta\left[\frac{9\Theta+6/\eta\tau}{6\Theta+8/\eta\tau} \right]   
\end{align}
Here, $\rho$ is the mass density, $\Theta=p/\rho c^2$, $p$ is the gas pressure, $\eta=m_{\rm el}/m_{\rm p}$, $m_{\rm p}$ is the mass of the proton, $\xi=n_{\rm p}/n_{\rm el}$ is the ratio of the number density of protons to the electron and $\tau=2-\xi+\xi/\eta$. The charge balance is maintained by the protons, electrons, and positrons. If $\xi=1$, it implies electron-proton plasma; on the other, $\xi=0$ implies electron-positron pair plasma. Any other $\xi$ between $0$---$1$, would imply electron-positron-proton plasma.}

\section{Results}

Numerical simulation of radiative acceleration of jets above an advective disc
has been carried out in one (1D) and two spatial dimensions (2D), starting with subsonic, sub-relativistic jets to supersonic, relativistic jets in detail by \cite{jdc22, jctt24}. The 1D jet structure is way simpler, and multidimensional simulations create a lot more complicated structures, primarily due to various instabilities like Kelvin-Helmholtz type instabilities. As a result, in 1D simulations, one could get impressive ($>50$) terminal Lorentz factors for pair-dominated flow \cite{jdc22}, but for 2D simulations, jets reach Lorentz factors up to $\sim 30$ \cite{jctt24}. 

\begin{figure}
	\includegraphics[width=12cm,height=5cm]{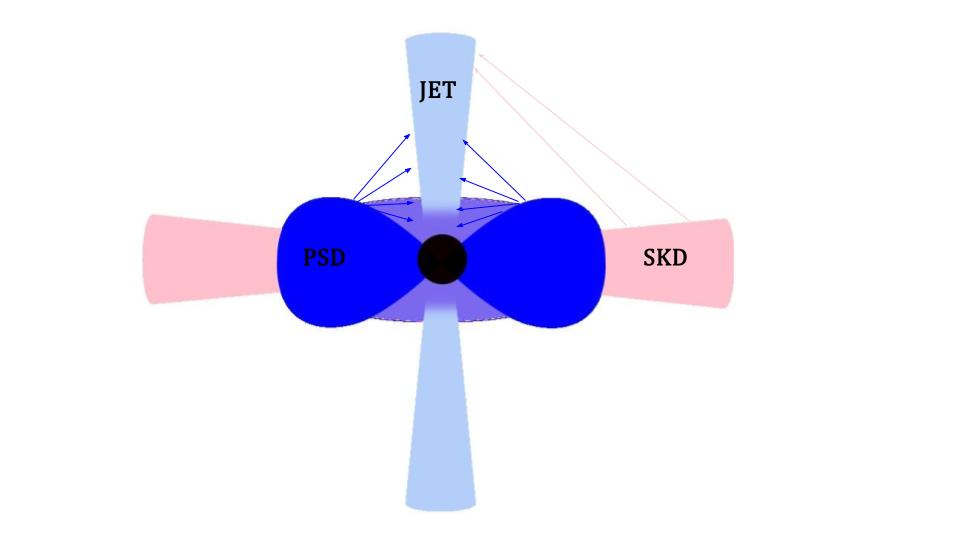}
  \caption{Cartoon diagram of the cross-section of the advective accretion disc and associated jets. The post-shock disc, or PSD, and the pre-shock sub-Keplerian disc, or SKD, are shown. The arrows represent the radiation impinging on the jet.}
  \label{fig:cartun}
\end{figure}

In this paper, we focus on comparing the radiatively driven jets from the advective disc, {the inner part of which are} geometrically thick, with that of discs that were geometrically slim {or thin}.
It may be noted that an advective-transonic disc \cite{f87,c89,c96} consists of matter falling onto a black hole through one or more sonic/critical points. In case matter becomes transonic at a larger distance from the black hole, it can go through a standing or a time-dependent shock before falling onto the black hole through the inner sonic point. It has been shown through various numerical simulations \cite{msc96,mrc96,lckr16,gk23} that such discs, when undergoing shocks, generate bipolar outflows from the post-shock region, which were proposed as precursors of astrophysical jets. 
Therefore, one can consider the shocks as the base of the jet. In Fig. \ref{fig:cartun}, we present a schematic diagram of the cross-section of disc-jet geometry. The post-shock disc (PSD), the sub-Keplerian disc (SKD), and the jet are shown. The arrows represent radiations from PSD and SKD interacting with the jets, which are flowing perpendicular to the disc plane.

\begin{figure}
	\includegraphics[width=\textwidth]{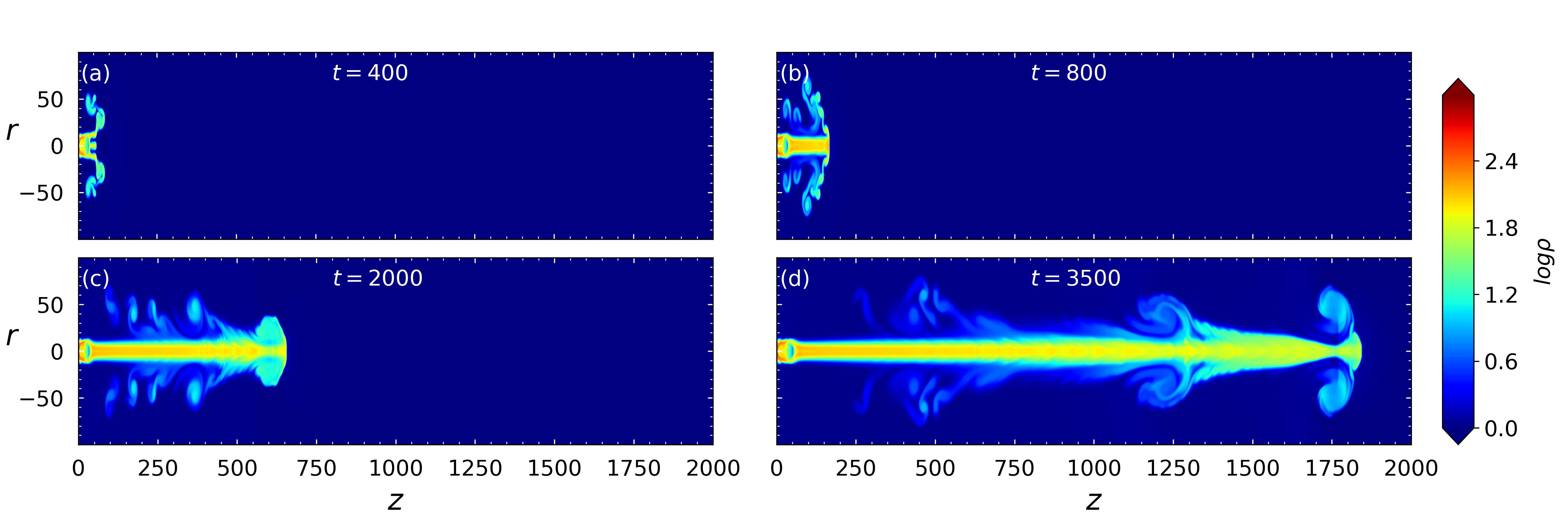}
    \caption{Contours of $log$ $\rho$ for an electron-positron jet driven by the radiation field of an accretion disc with $x_s=10.0,\,h_s=2.5x_s$. The disc luminosity is $l=0.2{L}_{Edd}$ and $\dot{m}=7.5\dot{M}_{Edd}$. The injection parameters for the jet beam are taken as $v_{\rm inj}=0.001,\,\Theta=2.0$ at $z=3.0$. The jet beam is 1000 times heavier than the ambient medium and $p_{\rm a}=p_{\rm j}/10.0$}
    \label{fig:thick_disc_xi0}
\end{figure}

The vertical height of PSD ($h_s$) can be geometrically thick ($h_s/x_s>1$ or can be geometrically thin ($h_s/x_s<1$), where $x_s$ is the radial extent of the PSD on the equatorial plane. 
In Fig. \ref{fig:thick_disc_xi0}, we plot the evolution of a jet of fluid composed of purely electron and positron {(i. e., $\xi=0$)}, powered by radiation from an accretion disc, in which the dimensions of PSD are given by $h_s/x_s=2.5$ and $x_s=10r_{\rm g}$. {The initial ratio of the ambient ($p_{\rm a}$) to the jet pressure ($p_{\rm j}$) is $0.1$.}
The jet material remains denser than the ambient medium, and the backflow creates significant Kelvin-Helmhotlz plumes. The radiation field is such that there is a significant back flow at around $z\sim 6 r_{\rm g}$; however, picks up the speed and is accelerated by radiation to about $0.8c$, followed by a strong shock at around $40 r_{\rm g}$ (bluish region on the jet spine) and a compression ratio of about $4$.
The terminal speed of this jet along the spine is $\sim 0.6 c$. 

\begin{figure}
	\includegraphics[width=\textwidth]{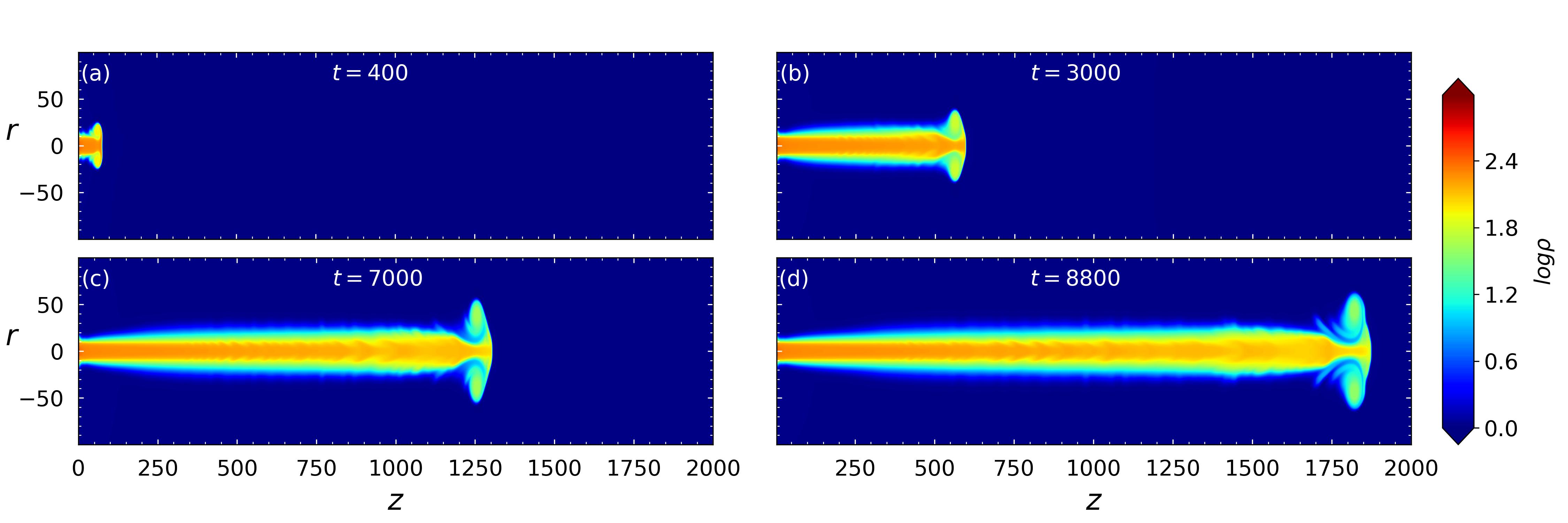}
    \caption{Contours of $log$ $\rho$ for an electron-proton jet. The accretion disc parameters and injection parameters at the jet base are the same as mentioned in Fig. \ref{fig:thick_disc_xi0}}
    \label{fig:thick_disc_xi1}
\end{figure}

In Fig. \ref{fig:thick_disc_xi1}, we plot the evolution of the density profile of an electron-proton jet but with the same jet injection and disc parameters of
Fig. \ref{fig:thick_disc_xi0}. Radiative acceleration of electron-proton medium is not a very efficient process. With the sub-Eddington disc luminosity, an electron-proton jet has less proportion of electrons, and therefore, radiation has a limited effect on the jet. {Near the base, the jet} expands due to thermal driving, reaching up to relatively high speeds $v\sim 0.45c$. {Thereafter}, the radiation drag {slows the jet down to speeds up to} $v\sim 0.25c$.

\begin{figure}
	\includegraphics[width=9cm,height=5cm]{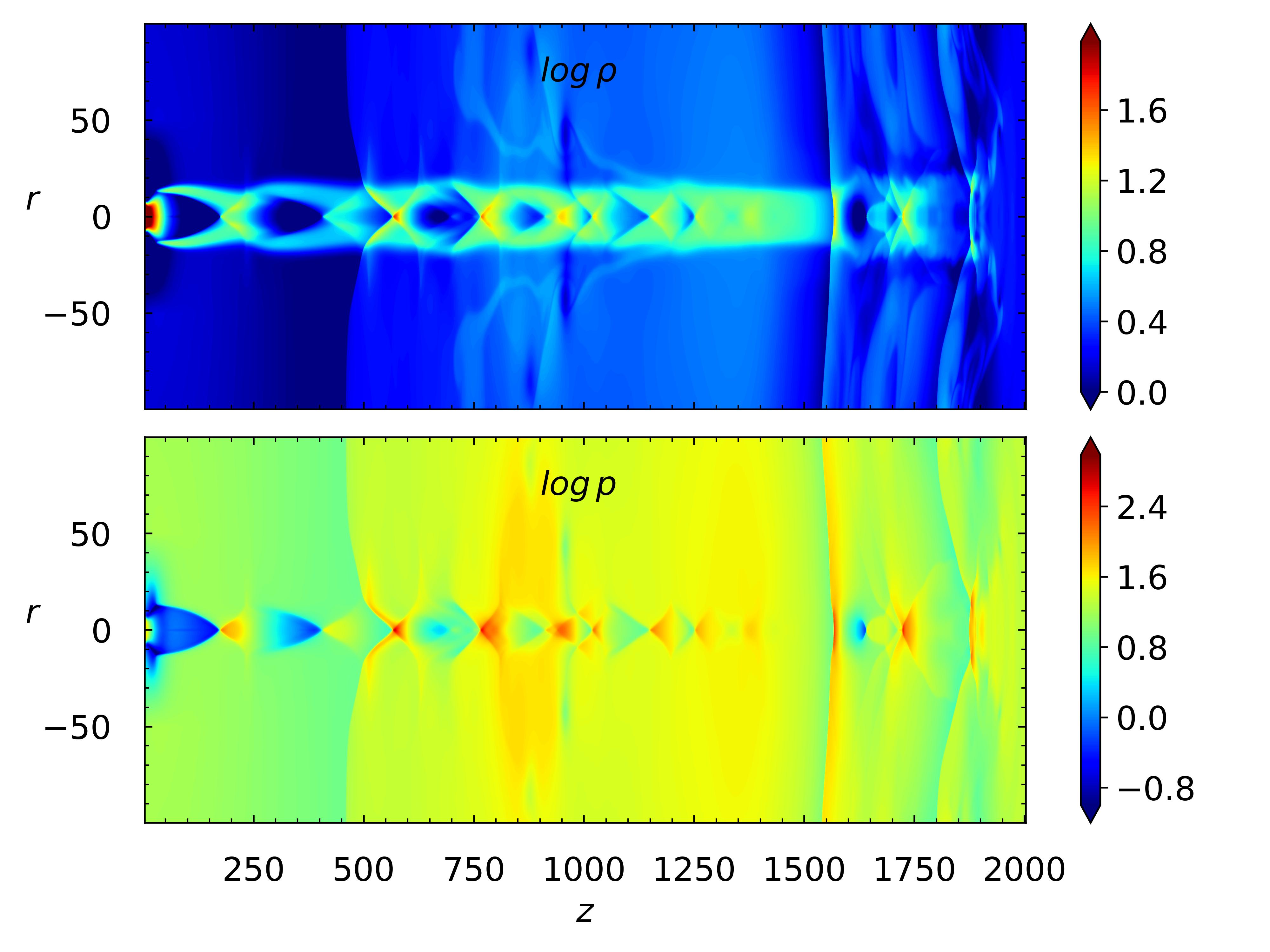}
    \caption{Contours of $log$ $\rho$ (left) and $log$ $P$ (right) in the $r-z$ plane for a pair dominated jet $\xi=0.001$. For this case $h_s=0.6(x_s-1.0)$. The disc luminosity is $l=0.35{L}_{\rm Edd}$ and $\dot{m}=9.5\dot{M}_{\rm Edd}$. The injection parameters for the jet beam are taken as $v_{\rm inj}=0.001,\,\Theta=0.1$ at $z=3.0$}
    \label{fig:thin_disk}
\end{figure}

{The traditional thick disc or the Polish doughnut \cite{pw80}, is without advection and is unstable. A} geometrically thick, advective disc {is stable, but the radiation field produced is such that}, the radiative acceleration is moderate. However, for an advective disc that is geometrically slim, the radiation field looks away from the disc and accelerates the jet.
In Fig. \ref{fig:thin_disk}, we plot the contours of $log$ $\rho$ (left panel) and $log$ $P$ (right panel){of a pair dominated jet, and is acted on by the radiation field around a slim advective disc (the injection values and other factors are mentioned in the caption).}
{Since} the jets are accelerated, there are many shocks in the jet spine, which causes the braided structure of shocks and rarefaction regions. The shocks also show up as enhancements in the pressure plot.

\begin{figure}
	\includegraphics[width=10cm,height=5cm]{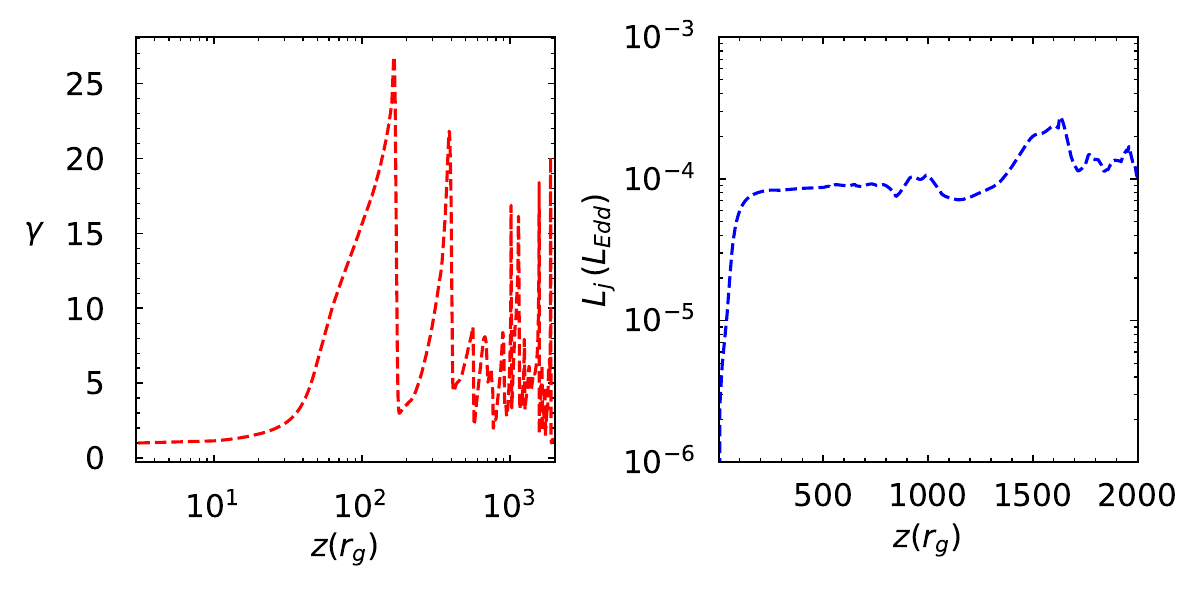}
    \caption{Variation of Lorentz factor $\gamma$ (left) and jet kinetic luminosity $L_j$ (right) as a function of $z$ for jet model presented in Fig. \ref{fig:thin_disk}}
    \label{fig:spine_vars}
\end{figure}

In Fig. \ref{fig:spine_vars} (left), we plot the bulk Lorentz factor along the axis of the jet for the case presented in the previous figure, and in the right panel, we plot the jet kinetic luminosity. The presence of many shocks implies that the jet speed profile along \textbf{its} spine will not be smooth. All the vertical jumps in the left panel represent shocks, and the spikes in $\gamma$ coincide with the rarefaction regions (dark blue regions in the braided structure of the jet) in the left panel of Fig. \ref{fig:thin_disk}. In these rarefaction regions, the Lorentz factor reaches $\gamma \gsim 20$!

\begin{figure}
	\begin{center}
	    \includegraphics[width=8cm,height=6.5cm]{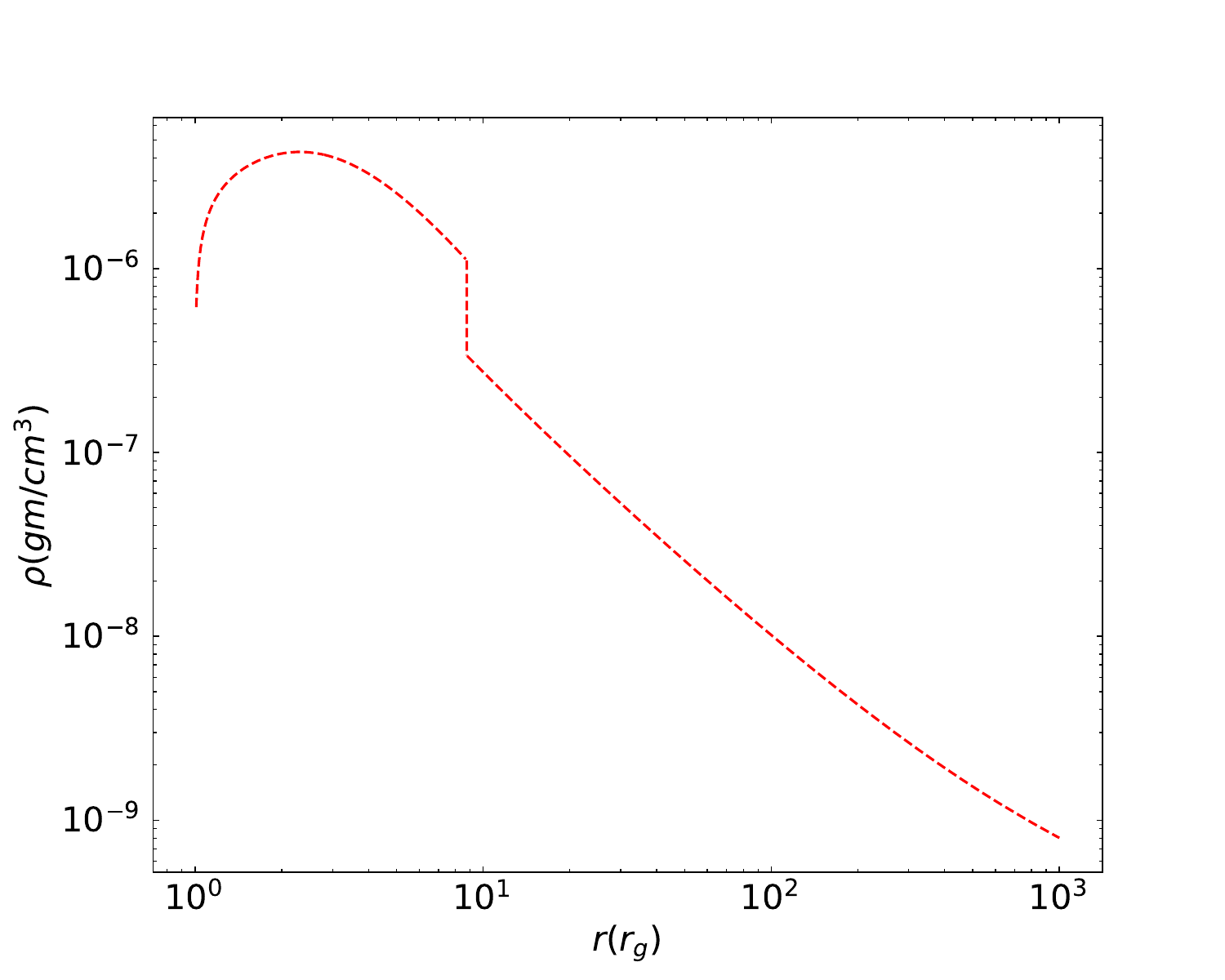}
	\end{center}
    \caption{Variation of mass density along the equatorial plane of the accretion disk for accretion rate $\dot{m}=10\dot{M}_{Edd}$, generalized specific energy $\epsilon= 1.0001$ and specific angular momentum at the horizon $\lambda_0=1.6755$, for a viscosity parameter $\alpha_{\rm v}=0.005$, in the presence of bremsstrahlung and synchrotron and a simplified inverse-Compton processes. Adopted from \cite{kc14}. All have been computed assuming a black hole mass $10M_\odot$.}
    \label{fig:acc_dens}
\end{figure}

\section{Discussions}

Radiative acceleration of jets, as a process, has never been favoured in the community, mainly on account of radiation drag effects, as has been discussed in the introduction \cite{sw81,i89,f96}. Radiation drag is effective when the source is extended, so all the arguments made in favour of dominant radiation drag involved either infinitely extended discs or the jet being studied in the funnel of a very thick torus. Later, a few simulations were done with jet generation from an initial thick tori and very high accretion rates \cite{sn15,ommk09}. The medium above the accreting tori becomes optically thick to slim.
{Therefore, photons emerging from the accreting torus of a classical thick disc undergoes many scatterings, thereby increasing the percentage of back scattered photons. This}
limits the scope of acceleration of the outflowing jets, resulting in mildly relativistic speeds. 

In the case of the advective discs, the shocked state is associated with the jet state. {It has been shown that shock in accretion can form, even in the} presence of significant advection and super-critical accretion rate, and yet {because of the presence of strong advection,} the disc densities are orders of magnitude lower, {than that in super critical accretion tori}. In Kumar \& Chattopadhyay (2014) \cite{kc14}, we presented shocked accretion solutions with super critical accretion rates and we reproduce one such solution in Fig. \ref{fig:acc_dens}. The most interesting part is that even for such high values of accretion rate, the maximum density turns out to be $\sim 10^{-6}$g cm$^{-3}$, which is orders of magnitude less
than typical thick disc solutions of similar accretion rates. As mentioned before, the presence of advection makes the advective discs more stable and of lower density. 
The presence of advection also ensures that from the post-shock region, only a fraction of inflowing matter can be redirected as outflows. This fact too, makes the outflows optically thin. As a result, the radiation can accelerate the plasma above an advective disc.

\section{Summary and conclusion} 
In this paper, we explored the acceleration scenario of jets by radiation from the advective disc. Simulations show that the advective discs can be geometrically thick or thin, and that creates different radiation fields. We show that for thick discs, even electron-positron jets can be accelerated up to Lorentz factors of few. Electron proton jets are not significantly affected by radiation. The pair-dominated flows from thin, advective discs can be accelerated to ultra-relativistic speeds, and the jets can produce a large number of shocks as well. If the accretion rate is more than a hundred Eddington rate, then the jet will become optically thick, and this method of radiative acceleration will become ineffective.

\begin{acknowledgement}
We thank the organizers of ISRA 2023. Our special thanks to Shubhrangshu Ghosh and Banibrata Mukhopadhyay. 
\end{acknowledgement}


\begin{thebibliography}{99.}%
% Use the following syntax and markup for your references if 
% the subject of your book is from the field 
% "Mathematics, Physics, Statistics, Computer Science"
%
% Contribution 
%\bibitem{SBbook} S. Banerjee, {\it Open Quantum Systems: Dynamics of Nonclassical Evolution} {Springer Singapore, 2018}.

\bibitem{sw81} M. {Sikora}, D. B. {Wilson}, MNRAS, {\bf 197}, 529 (1981)  

\bibitem{i89} V. Icke, A\&A, {\bf 216}, 294 (1989)

\bibitem{f96} J. Fukue, PASJ, {\bf 48}, 631 (1996)

\bibitem{f99} J. Fukue, PASJ, {\bf 51}, 425 (1999)

\bibitem{f05} J. Fukue, PASJ, {\bf 57}, 691 (2005)

\bibitem{fth01} J. Fukue, M. Tojyo, Y. Hirai, PASJ, {\bf 53}, 555 (2001)

\bibitem{eg85} G. E. Eggum, J. I. Katz, ApJ, {\bf 298}, L41 (1985)

\bibitem{sn15} A. Sadowski, R. Narayan, MNRAS,
{\bf 453}, 3213 (2015)

\bibitem{ommk09} K. Oshuga, S. Mineshige, M. Mori, Y. Kato, PASJ, {\bf 61}, L7 (2009)

\bibitem{cc00a} I. Chattopadhyay, S. K. Chakrabarti, Int. Journ. Mod. Phys. D, {\bf 9}, 57 (2000)

\bibitem{cc00b} I. Chattopadhyay, S. K. Chakrabarti, Int. Journ. Mod. Phys. D, {\bf 9}, 717  (2000)

\bibitem{cc02} I. Chattopadhyay, S. K. Chakrabarti, MNRAS, {\bf 333}, 454 (2002)

\bibitem{cdc04} I. Chattopadhyay, S. Das, S. K. Chakrabarti, MNRAS, {\bf 348}, 846 (2004)

\bibitem{c05} I. Chattopadhyay, MNRAS, {\bf 356}, 145 (2005)

\bibitem{vkmc15} M. K. Vyas, R. Kumar, S. Mandal, I. Chattopadhyay, MNRAS, {\bf 453}, 2992 (2015)

\bibitem{jdc22} R. K. Joshi, S. Debnath, I. Chattopadhyay, ApJ, {\bf 933}, 75 (2022)

\bibitem{jctt24} R. K. Joshi, I. Chattopadhyay, A. Tsokaros, P. K. Tripathi, ApJ, \textbf{971}, 13 (2024) %, DOI 10.3847/1538-4357/ad54cO

\bibitem{msc96} D. Molteni, H. Sponholz, S. K. Chakrabarti, ApJ, {\bf 457}, 805 (1996)

\bibitem{mrc96} D. Molteni, D. Ryu, S. K. Chakrabarti, ApJ, {\bf 470}, 460 (1996)

\bibitem{lckr16} S. J. Lee, I. Chattopadhyay, R. Kumar, D. Ryu, ApJ, {\bf 831}, 33 (2016)

\bibitem{dc14} S. Das, I. Chattopadhyay, A. Nandi, D. Molteni, MNRAS, {\bf 442}, 251 (2014)

\bibitem{rcc06} D. Ryu, I. Chattopadhyay, E. Choi, ApJS, {\bf 166}, 410 (2006)

\bibitem{pw80} B. Pacz\'ynsky, P. J. Wiita, A\&A, {\bf 88}, 23 (1980)

\bibitem{cr09} I. Chattopadhyay, D. Ryu, ApJ, {\bf 694}, 492 (2009) 

\bibitem{f87} J. Fukue, PASJ, {\bf 39}, 309 (1987)

\bibitem{c89} S. K. Chakrabarti, Apj, {\bf 347}, 365 (1989)

\bibitem{c96} S. K. Chakrabarti, ApJ, {\bf 464}, 664 (1996)

\bibitem{ct95} S. K. Chakrabarti, L. Titarchuk, ApJ, {\bf 455}, 623 (1995)

\bibitem{gk23} S. Garain, J. Kim, MNRAS, {\bf 519}, 4550 (2023)

\bibitem{vc19} M. K. Vyas, I. Chattopadhyay, MNRAS, {\bf 482}, 4203 (2019)

\bibitem{kc14} R. Kumar, I. Chattopadhyay, MNRAS, {\bf 443}, 3444 (2014)


%\bibitem{wl73} W. H. Louisell, {\it Quantum Statistical Properties of Radiation} (John Wiley and Sons, 1973). 
%\bibitem{cl83} A. O. Caldeira and A. J. Leggett, Phy, sica A {\bf 121}587 (1983). 
%\bibitem{wz93} W. H. Zurek, Phys. Today {\bf 44}, 36 (1991); 
%Prog. Theor. Phys. {\bf 87}, 281 (1993).
%\bibitem{turch} Q. A. Turchette, C. J. Myatt, B. E. %King, C. A. Sackett, {\it et al.}, Phys. Rev. A {\bf 62},
%053807 (2000).
%\bibitem{myatt}   C.  J.   Myatt,  B.   E.  King,   %Q.   A.  Turchette,
%C. A. Sackett, {\it et al.}, Nature  {\bf 403}, 269 (2000).
%\bibitem{haroche} M. Brune, {\it et al.}, Phys. Rev. Lett. {\bf 77}, 4887 (1996).
%\bibitem{kim} C. Giunti and C. W. Kim, Fundamentals of Neutrino Physics and Astrophysics (Oxford, 2007; online edn, Oxford Academic, 1 Jan. 2010).
%\bibitem{Dune} V. A. Kudryavtsev and for the DUNE Collaboration, J. Phys.: Conf. Ser. {\bf 718}, 062032 (2016).
% \bibitem{bg03} S. Banerjee and R. Ghosh, Phys. Rev. E {\bf 
% 67}, 056120 (2003). 
% \bibitem{hpz92} B.L. Hu, J.P. Paz and Y. Zhang, Phys. Rev. D {\bf 45}, 2843
% (1992); {\it ibid.} {\bf 47}, 1576 (1993); B.L. Hu and A. Matacz, Phys. Rev. D.
% {\bf 49}, 6612 (1994).
% \bibitem{ps91} J. P. Paz and S. Sinha, Phys. Rev. D {\bf 44}, 1038 (1991);
% {\bf 45}, 2823 (1992).
% \bibitem{ah93} A. Anderson and J.J. Halliwell, Phys. Rev. D {\bf 48},
% 2753 (1993).
% \bibitem{gupta84} R. K. Gupta, M. M unchow, A. Sandulescu and W. Scheid, J. Phys. G. {\bf 10}, 209 (1984).
% \bibitem{isar93} A. Isar, A. Sandulescu and W. Scheid, J. Math. Phys. {\bf 34}, 3887 (1993).
% \bibitem{gsa74} G. S. Agarwal, {\it Quantum Statistical Theories of Spontaneous Emission and their Relation to other Approaches}
% (Springer Tracts in Modern Physics, 1974).
% \bibitem{bsrik07} R. Srikanth and S. Banerjee, Phys. Lett. A {\bf 367}, 295 (2007); Phys. Rev. A {\bf 77}, 012318 (2008).
% \bibitem{land61} R. Landauer, IBM J Res. Dev. {\bf 5}, 183 (1961).
% \bibitem{ben82} C. H. Bennett, Int. J. Theor. Phys. {\bf 21}, 905 (1982).
% \bibitem{butt92} M. Buttiker, Jr. of Phys. Cond. Matt. {\bf 5}, 9361 (1993).
% \bibitem{meir92} Y. Meir and N. S. Wingreen, Phys. Rev. Lett. {\bf 68}, 2512 (1992).
% \bibitem{datta96} S. Datta, {\it Electronic Transport in Mesoscopic Systems} (Cambridge University Press, Cambridge, 1995).
% \bibitem{imry97} Y. Imry, {\it Introduction to Mesoscopic Physics} (Oxford University Press, 2002).
% \bibitem{brs09} S. Banerjee, V. Ravishankar and R. Srikanth, Ann. of Phys. (N.Y.) {\bf 325}, 816 (2010).
% \bibitem{kram40} H. A. Kramers, Physica {\bf 7}, 284 (1940).
% \bibitem{leg87} A. J. Leggett, {\it et al.}, Rev. Mod. Phys. {\bf 59}, 1 (1987).
% \bibitem{ambe82} V. Ambegaokar, U. Eckern, and G. Schon, Phys. Rev. Lett. {\bf 48}, 1745 (1982).
% \bibitem{fiszwe85} M. P.A. Fisher and W. Zwerger, Phys. Rev. B {\bf 32}, 6190 (1985).
%\bibitem{brag75} V. B. Braginsky, Yu. I. Vorontsov and K. S. Thorne, Science {\bf 209}, 547 (1980).
%\bibitem{cave80} C. M. Caves, {\it et al.}, Rev. Mod. Phys. {\bf 52}, 341 (1980). 
%\bibitem{gsi88} H. Grabert, P. Schramm and G. L. Ingold, 
%Phys. Rep. {\bf 168}, 115 (1988). 
% \bibitem{gango01} G. Gangopadhyay, M. S. Kumar and S. Dattagupta, J. Phys. A: Math. Gen. {\bf 34}, 5485 (2001).
% \bibitem{bg07} S. Banerjee and R. Ghosh, J. Phys. A: Math. Gen. {\bf 40}, 13735 (2007).
%\bibitem{fv63} R. P. Feynman and F. L. Vernon, Ann. Phys. (N.Y.) {\bf 24}, 118 (1963).
%\bibitem{ha85} V. Hakim and V. Ambegaokar, Phys. Rev. A {\bf 32}, 423 (1985).
% \bibitem{unruh95} W. G. Unruh, Phys. Rev. A {\bf 51}, 992 (1995).
% \bibitem{palma96} G. M. Palma, K-A Suominen and A. K. Ekert,  Proc. R. Soc. Lond. A {\bf 452}, 567 (1996).
% \bibitem{di95} D. P. DiVincenzo, Phys. Rev. A {\bf 51}, 1015 (1995).
%\bibitem{nc00} M. Nielsen and I. Chuang, {\it Quantum Computation and Quantum Information} (Cambridge University Press, Cambridge, 2000).
%\bibitem{sting55} W. F. Stinespring, {\it Positive Functions on C*-algebras} (Proceedings of the American Mathematical Society, 211–216, (1955)).
%\bibitem{sud61} E. C. G. Sudarshan, {\it et al.}, Phys. Rev. {\bf 121}, 920 (1961).
%\bibitem{kr83} K. Kraus, {\it States, Effects and Operations: Fundamental Notions of Quantum Theory} (Springer Verlag, 1983).
%\bibitem{Horodecki_1996} R. Horodecki, M. Horodecki, and P. Horodecki, Physics Letters A, {\bf 222}, 21-25 (1996).
%\bibitem{Wootters_1998} W. K. Wootters, Phys. Rev. Lett. {\bf 80}, 2245 (1998).
%\bibitem{Luo_2008} S. Luo, Phys. Rev. A {\bf 77}, 022301 (2008).
%\bibitem{Zurek_2001} H. Ollivier and W. H. Zurek, Phys. Rev. Lett. {\bf 88}, 017901 (2001). 
%\bibitem{Vedral_2001} L. Henderson and V. Vedral, J. Phys. A.: Math. Gen. \textbf{34}, 6899 (2001).
%\bibitem{LG_1985} A. J. Leggett and A. Garg, Phys. Rev. Lett. \textbf{54}, 857 (1985).
%\bibitem{Brukner_2010} B. Dakić, V. Vedral, and Č. Brukner
%Phys. Rev. Lett. \textbf{105}, 190502 (2010).
%\bibitem{Blasone_2008} M. Blasone, F. Dell'Anno, S. De Siena, M. Di Mauro, and F. Illuminati, Phys. Rev. D \textbf{77}, 096002 (2008)
%\bibitem{Blasone_2009} M. Blasone, F. Dell'Anno, S. De Siena, and F. Illuminati, Europhys. Lett. {\bf 85}, 50002 (2009).
%\bibitem{SB_2016} A. K. Alok, S. Banerjee, S. Uma Sankar, Nucl. Phys. B \textbf{909}, 65 (2016).
%\bibitem{SB_2015} S. Banerjee, A. K. Alok, R. Srikanth, B. C. Hiesmayr, Eur. Phys. J. C {\bf 75}, 487 (2015).
%\bibitem{SB_2019} J. Naikoo, A. K. Alok, S. Banerjee, and S. Uma Sankar, Phys. Rev. D {\bf 99}, 095001 (2019).
%\bibitem{SB_2020} J. Naikoo, S. Kumari, S. Banerjee, A. K. Pan, J. Phys. G {\bf 47}, 095004 (2020).
%\bibitem{SB_2018} K. Dixit, J. Naikoo, S. Banerjee, A. K. Alok, Eur. Phys. J. C {\bf 78}, 914 (2018).
%\bibitem{upcoming} S. Bouri, A. K. Jha, S. Banerjee, B. Mukhopadhyay, in progress. 
%\bibitem{SB_BM_2018} K. Dixit, J. Naikoo, B. Mukhopadhyay, S. Banerjee, Phys. Rev. D {\bf 100}, 055021 (2019).
%\bibitem{AK_SB_2013} A. K. Alok, S. Banerjee, Phys. Rev. D {\bf 88}, 094013 (2013).
%\bibitem{AK_SB_MK_2016} S. Banerjee,  A. K. Alok, R. MacKenzie, Eur. Phys. J Plus {\bf 131}, 129 (2016).
%\bibitem{PDG_2014} Particle Data Group Collaboration (K. A. Olive \textit{et. al.}), Chin. Phys. C {\bf 38}, 090001 (2014).
%\bibitem{HFAG_2012} Heavy Favor Averaging Group Collaboration (Y. Amhis {\it et. al.}), arXiv:1907.1158 (2012).
%\bibitem{KLOE_2006} KLOE Collaboration (F. Ambrosino {\it et. al.}), Phys. Lett. B {\bf 642}, 315 (2006).
%\bibitem{Grimus_2001} R. A. Bertlmann and W. Grimus, Phys. Rev. D {\bf 64}, 056004 (2001).
%\bibitem{AK_SB_SU_2015} A. K. Alok, S. Banerjee, S. Uma Sankar, Phys. Lett. B {\bf 749}, 94 (2015).
%\bibitem{JK_SB_2018} J. Naikoo, A. K. Alok, and S. Banerjee, Phys. Rev. D {\bf 97}, 053008 (2018). 
%
% \bibitem{panch56} S. Pancharatnam, Proc. Indian Acad. Sci., Sect. A {\bf 44}, 247 (1956).
% \bibitem{ber84} M. V. Berry, Proc. R. Soc. London, Ser. A {\bf 392}, 45 (1984).
% \bibitem{sim83} B. Simon, Phys. Rev. Lett. {\bf 51}, 2167 (1983).
% \bibitem{aa87} Y. Aharonov and J. Anandan, Phys. Rev. Lett. {\bf 58}, 1593 (1987).
% \bibitem{sb88} J. Samuel and R. Bhandari, Phys. Rev. Lett. {\bf 60}, 2339 (1988).
% \bibitem{ms93} N. Mukunda and R. Simon, Ann. Phys. (N.Y.) {\bf 228}, 205 (1993).
% \bibitem{sj00}E. Sjoqvist {\it et al.}, Phys. Rev. Lett. {\bf 85}, 2845 (2000).
% \bibitem{tong04} D. M. Tong, E. Sjoqvist, L. C. Kwek and C. H. Oh, Phys. Rev. Lett. {\bf 93}, 080405 (2004).
% \bibitem{falci0} G. Falci, {\it et al.}, Nature (London) {\bf 407}, 355 (2000).
% \bibitem{naka99} Y. Nakamura, Yu. A. Pashkin and J. S. Tsai, Nature (London) {\bf 398}, 786 (1999).
% \bibitem{whit03} R. S. Whitney and Y. Gefen, Phys. Rev. Lett. {\bf 90}, 
% 190402 (2003); R. S. Whitney, Y. Makhlin, A. Shnirman and Y. Gefen, ibid.
% {\bf 94}, 070407 (2005).
% \bibitem{bsgp07} R. Srikanth and S. Banerjee, Phys. Rev. A {\bf 77}, 012318 (2008).
% \bibitem{crysw} N. Srinatha, {\it et al.}, Quantum Inf Process {\bf 13}, 59 (2014).
% \bibitem{sh95} P. W. Shor, SIAM J. Sci. Statist. Comput. {\bf 26}, 1484 (1997);
% eprint quant-ph/9508027; L. K. Grover, Phys. Rev. Lett. {\bf 79}, 325 (1997).
% \bibitem{wz82} W. K. Wooters and W. H. Zurek, Nature {\bf 299}, 802 (1982).
% \bibitem{di82} D. Dieks, Physics Letters A {\bf 92}, 271 (1982).
% \bibitem{pb00} A. K. Pati and S. Braunstein, Nature {\bf 404}, 164 (2000).
% \bibitem{cs96}A. R. Calderbank and P. W. Shor, Phys. Rev. A {\bf 54}, 1098 (1996);
% A. Steane, Proc. Roy. Soc., London, Ser. A {\bf 452}, 2551 (1996).
% \bibitem{vl98} L. Viola and S. Lloyd, Phys. Rev. A {\bf 58}, 2733 (1998);
% D. Vitali and P. Tombesi, Phys. Rev. A {\bf 65}, 012305 (2001).
% \bibitem{lc98} D. A. Lidar, I. L. Chuang and K. B. Whaley,
% Phys. Rev. Lett. {\bf 81}, 2594 (1998).


% \bibitem{weiss} U. Weiss, {\it Quantum Dissipative Systems} (World Scientific, 2012).
% \bibitem{Breuer} H. Breuer and F. Petruccione,  {\it Open Quantum Systems} (Oxford University Press,  2002). 

% \bibitem{science-contrib} Broy, M.: Software engineering --- from auxiliary to key technologies. In: Broy, M., Dener, E. (eds.) Software Pioneers, pp. 10-13. Springer, Heidelberg (2002)
% %
% % Online Document
% \bibitem{science-online} Dod, J.: Effective substances. In: The Dictionary of Substances and Their Effects. Royal Society of Chemistry (1999) Available via DIALOG. \\
% \url{http://www.rsc.org/dose/title of subordinate document. Cited 15 Jan 1999}
% %
% % Monograph
% \bibitem{science-mono} Geddes, K.O., Czapor, S.R., Labahn, G.: Algorithms for Computer Algebra. Kluwer, Boston (1992) 
% %
% % Journal article
% \bibitem{science-journal} Hamburger, C.: Quasimonotonicity, regularity and duality for nonlinear systems of partial differential equations. Ann. Mat. Pura. Appl. \textbf{169}, 321--354 (1995)
% %
% % Journal article by DOI
% \bibitem{science-DOI} Slifka, M.K., Whitton, J.L.: Clinical implications of dysregulated cytokine production. J. Mol. Med. (2000) doi: 10.1007/s001090000086 
% %
% %\bigskip

% % Use the following (APS) syntax and markup for your references if 
% % the subject of your book is from the field 
% % "Mathematics, Physics, Statistics, Computer Science"
% %
% % Online Document
% \bibitem{phys-online} J. Dod, in \textit{The Dictionary of Substances and Their Effects}, Royal Society of Chemistry. (Available via DIALOG, 1999), 
% \url{http://www.rsc.org/dose/title of subordinate document. Cited 15 Jan 1999}
% %
% % Monograph
% \bibitem{phys-mono} H. Ibach, H. L\"uth, \textit{Solid-State Physics}, 2nd edn. (Springer, New York, 1996), pp. 45-56 
% %
% % Journal article
% \bibitem{phys-journal} S. Preuss, A. Demchuk Jr., M. Stuke, Appl. Phys. A \textbf{61}
% %
% % Journal article by DOI
% \bibitem{phys-DOI} M.K. Slifka, J.L. Whitton, J. Mol. Med., doi: 10.1007/s001090000086
% %
% % Contribution 
% \bibitem{phys-contrib} S.E. Smith, in \textit{Neuromuscular Junction}, ed. by E. Zaimis. Handbook of Experimental Pharmacology, vol 42 (Springer, Heidelberg, 1976), p. 593
% %
% %\bigskip
% %
% % Use the following syntax and markup for your references if 
% % the subject of your book is from the field 
% % "Psychology, Social Sciences"
% %
% %
% % Monograph
% \bibitem{psysoc-mono} Calfee, R.~C., \& Valencia, R.~R. (1991). \textit{APA guide to preparing manuscripts for journal publication.} Washington, DC: American Psychological Association.
% %
% % Online Document
% \bibitem{psysoc-online} Dod, J. (1999). Effective substances. In: The dictionary of substances and their effects. Royal Society of Chemistry. Available via DIALOG. \\
% \url{http://www.rsc.org/dose/Effective substances.} Cited 15 Jan 1999.
% %
% % Journal article
% \bibitem{psysoc-journal} Harris, M., Karper, E., Stacks, G., Hoffman, D., DeNiro, R., Cruz, P., et al. (2001). Writing labs and the Hollywood connection. \textit{J Film} Writing, 44(3), 213--245.
% %
% % Contribution 
% \bibitem{psysoc-contrib} O'Neil, J.~M., \& Egan, J. (1992). Men's and women's gender role journeys: Metaphor for healing, transition, and transformation. In B.~R. Wainrig (Ed.), \textit{Gender issues across the life cycle} (pp. 107--123). New York: Springer.
% %
% % Journal article by DOI
% \bibitem{psysoc-DOI}Kreger, M., Brindis, C.D., Manuel, D.M., Sassoubre, L. (2007). Lessons learned in systems change initiatives: benchmarks and indicators. \textit{American Journal of Community Psychology}, doi: 10.1007/s10464-007-9108-14.
% %
% %
% % Use the following syntax and markup for your references if 
% % the subject of your book is from the field 
% % "Humanities, Linguistics, Philosophy"
% %
% %\bigskip
% %
% % Journal article
% \bibitem{humlinphil-journal} Alber John, Daniel C. O'Connell, and Sabine Kowal. 2002. Personal perspective in TV interviews. \textit{Pragmatics} 12:257--271
% %
% % Contribution 
% \bibitem{humlinphil-contrib} Cameron, Deborah. 1997. Theoretical debates in feminist linguistics: Questions of sex and gender. In \textit{Gender and discourse}, ed. Ruth Wodak, 99--119. London: Sage Publications.
% %
% % Monograph
% \bibitem{humlinphil-mono} Cameron, Deborah. 1985. \textit{Feminism and linguistic theory.} New York: St. Martin's Press.
% %
% % Online Document
% \bibitem{humlinphil-online} Dod, Jake. 1999. Effective substances. In: The dictionary of substances and their effects. Royal Society of Chemistry. Available via DIALOG. \\
% http://www.rsc.org/dose/title of subordinate document. Cited 15 Jan 1999
% %
% % Journal article by DOI
% \bibitem{humlinphil-DOI} Suleiman, Camelia, Daniel C. O'Connell, and Sabine Kowal. 2002. `If you and I, if we, in this later day, lose that sacred fire...': Perspective in political interviews. \textit{Journal of Psycholinguistic Research}. doi: 10.1023/A:1015592129296.
% %
% %
% %
% %\bigskip
% %
% %
% % Use the following syntax and markup for your references if 
% % the subject of your book is from the field 
% % "Computer Science, Economics, Engineering, Geosciences, Life Sciences"
% %
% %
% % Contribution 
% \bibitem{basic-contrib} Brown B, Aaron M (2001) The politics of nature. In: Smith J (ed) The rise of modern genomics, 3rd edn. Wiley, New York 
% %
% % Online Document
% \bibitem{basic-online} Dod J (1999) Effective Substances. In: The dictionary of substances and their effects. Royal Society of Chemistry. Available via DIALOG. \\
% \url{http://www.rsc.org/dose/title of subordinate document. Cited 15 Jan 1999}
% %
% % Journal article by DOI
% \bibitem{basic-DOI} Slifka MK, Whitton JL (2000) Clinical implications of dysregulated cytokine production. J Mol Med, doi: 10.1007/s001090000086
% %
% % Journal article
% \bibitem{basic-journal} Smith J, Jones M Jr, Houghton L et al (1999) Future of health insurance. N Engl J Med 965:325--329
% %
% % Monograph
% \bibitem{basic-mono} South J, Blass B (2001) The future of modern genomics. Blackwell, London 
%
\end{thebibliography}
\end{document}